\documentclass[12pt]{article}
\usepackage{epsfig}
\usepackage{amsmath}
\usepackage{amssymb}
\usepackage{color}
\def\Color#1{\color[named]{#1}}
\def\nl{\nonumber\\}
\newcommand{\GeV}{\unskip\,\mathrm{GeV}}
\def\mathswitchr#1{\relax\ifmmode{\mathrm{#1}}\else$\mathrm{#1}$\fi}

\newcommand{\Pd}{\mathswitchr d}
\newcommand{\Pu}{\mathswitchr u}
\newcommand{\Ps}{\mathswitchr s}
\newcommand{\Pc}{\mathswitchr c}

\newcommand\pubnumber{UTHEP-00-1001}
\newcommand\pubdate{\today}
\newcommand\hepnumber{hep-ph/0101261}

\def\csumb{ \Color{Green}$^a$Department of Physics and Astronomy\\
University of Tennessee, Knoxville, TN 37996-1200, USA\\
$^b$SLAC, Stanford University, Stanford, California 94309, USA,\\
$^c$CERN, Theory Division, CH-1211 Geneva 23, Switzerland,\\
$^d$DESY-Zeuthen, Theory Division, D-15738 Zeuthen, Germany,\\
$^e$Institute of Nuclear Physics,
  ul. Kawiory 26a, 30-055 Cracow, Poland,\\
$^f$Institute of Computer Science, Jagellonian University,\\
        ul. Nawojki 11, 30-072 Cracow, Poland}
\def\support{\footnote{\Color{Red}Work partly supported by the Maria Sk\l{}odowska-Curie
Joint Fund II PAA/DOE-97-316, the European Commission
5-th Framework contract HPRN-CT-2000-00149, and
the US Department of Energy Contracts  DE-FG05-91ER40627
and   DE-AC03-76ER00515.}} 

\def\Title#1{\begin{center} {\Large\bf #1 } \end{center}}
\def\Author#1{\begin{center}{ \sc #1} \end{center}}
\def\Address#1{\begin{center}{ \it #1} \end{center}}

\newcommand\pubblock{\rightline{\begin{tabular}{l} \pubnumber\\
         \pubdate\\ \hepnumber \end{tabular}}}
\newenvironment{Abstract}{\begin{quotation}  }{\end{quotation}}
\newenvironment{Presented}{\begin{quotation} \begin{center} 
             Presented at the\end{center}
      \begin{center}\begin{large}}{\end{large}\end{center} \end{quotation}}
\def\Acknowledgments{\bigskip  \bigskip \begin{center}
          \large\bf Acknowledgments\end{center}}

\makeatletter
\def\section{\@startsection{section}{0}{\z@}{5.5ex plus .5ex minus
 1.5ex}{2.3ex plus .2ex}{\large\bf}}
\def\subsection{\@startsection{subsection}{1}{\z@}{3.5ex plus .5ex minus
 1.5ex}{1.3ex plus .2ex}{\normalsize\bf}}
\def\subsubsection{\@startsection{subsubsection}{2}{\z@}{-3.5ex plus
-1ex minus  -.2ex}{2.3ex plus .2ex}{\normalsize\sl}}

\renewcommand{\@makecaption}[2]{%
   \vskip 10pt
   \setbox\@tempboxa\hbox{\small #1: #2}
   \ifdim \wd\@tempboxa >\hsize     % IF longer than one line:
       \small #1: #2\par          %   THEN set as ordinary paragraph.
     \else                        %   ELSE  center.
       \hbox to\hsize{\hfil\box\@tempboxa\hfil}
   \fi}

 \def\citenum#1{{\def\@cite##1##2{##1}\cite{#1}}}
\newcount\@tempcntc
\def\@citex[#1]#2{\if@filesw\immediate\write\@auxout{\string\citation{#2}}\fi
  \@tempcnta\z@\@tempcntb\m@ne\def\@citea{}\@cite{\@for\@citeb:=#2\do
    {\@ifundefined
       {b@\@citeb}{\@citeo\@tempcntb\m@ne\@citea\def\@citea{,}{\bf ?}\@warning
       {Citation `\@citeb' on page \thepage \space undefined}}%
    {\setbox\z@\hbox{\global\@tempcntc0\csname b@\@citeb\endcsname\relax}%
     \ifnum\@tempcntc=\z@ \@citeo\@tempcntb\m@ne
       \@citea\def\@citea{,}\hbox{\csname b@\@citeb\endcsname}%
     \else
      \advance\@tempcntb\@ne
      \ifnum\@tempcntb=\@tempcntc
      \else\advance\@tempcntb\m@ne\@citeo
      \@tempcnta\@tempcntc\@tempcntb\@tempcntc\fi\fi}}\@citeo}{#1}}
\def\@citeo{\ifnum\@tempcnta>\@tempcntb\else\@citea\def\@citea{,}%
  \ifnum\@tempcnta=\@tempcntb\the\@tempcnta\else
  {\advance\@tempcnta\@ne\ifnum\@tempcnta=\@tempcntb \else\def\@citea{--}\fi
    \advance\@tempcnta\m@ne\the\@tempcnta\@citea\the\@tempcntb}\fi\fi}
\makeatother

\def\beq{\begin{equation}}
\def\eeq#1{\label{#1}\end{equation}}
\def\eeqn{\end{equation}}

\newenvironment{Eqnarray}%
   {\arraycolsep 0.14em\begin{eqnarray}}{\end{eqnarray}}
\def\beqa{\begin{Eqnarray}}
\def\eeqa#1{\label{#1}\end{Eqnarray}}
\def\eeqan{\end{Eqnarray}}

\let\bar=\overbar

\def\half{\frac{1}{2}}

\def\Dslash{\not{\hbox{\kern-4pt $D$}}}
\def\dslash{\not{\hbox{\kern-2pt $\del$}}}

\def\msb{{\bar{\ssstyle M \kern -1pt S}}}

\def\lsim{\mathrel{\raise.3ex\hbox{$<$\kern-.75em\lower1ex\hbox{$\sim$}}}}
\def\gsim{\mathrel{\raise.3ex\hbox{$>$\kern-.75em\lower1ex\hbox{$\sim$}}}}

\begin{document}
\begin{titlepage}
\pubblock

\vfill
\def\thefootnote{\fnsymbol{footnote}}
\Title{\Color{Maroon}Precision Predictions for (Un)Stable\\[5pt] $WW/4f$
Production in $e^+e^-$ Annihilation:\\[5pt] YFSWW3/KoralW-1.42/YFSZZ\support}
\vfill
\Author{B.F.L. Ward$^{a,b,c}$, S. Jadach$^{c,d,e}$,
W. P\l{}aczek$^{c,f}$, M. Skrzypek$^{c,e}$ \\[5pt] and Z. W\c{a}s$^{c,e}$}
\Address{\csumb}
\vfill
\begin{Abstract}
{\Color{Blue}We present precision calculations of the processes 
$e^+e^-\rightarrow 4$-fermions in which the double resonant $W^+W^-$
and $ZZ$
intermediate states occur. Referring to these latter intermediate
states as the 'signal processes', we show that, by using the
YFS Monte Carlo event generators {\Color{Maroon}YFSWW3-1.14 and KoralW-1.42}
in an appropriate combination, we achieve a physical precision
on the $WW$ signal process, as isolated with LEP2 MC Workshop cuts,
{\Color{Green}below 0.5\%}. We stress the full gauge invariance of our calculations
and we compare our results with those of other authors where appropriate.
In particular, sample Monte Carlo data are explicitly illustrated
and compared with the results of the program 
{\Color{Maroon}RacoonWW} of Denner {\it et al.}.
In this way, we cross check that the total (physical$\oplus$technical) 
precision tag for the $WW$
signal process cross section is {\Color{Green}0.4\% for 200 GeV}, for example.
Results are also given for 500 GeV with an eye toward the LC.
For the analogous $ZZ$ case, we cross check that our {\Color{Maroon}YFSZZ}
calculation yields a total precision tag of {\Color{Red}2\%}, when it is
compared to
the results of {\Color{Maroon}ZZTO} and {\Color{Maroon}GENTLE} of 
Passarino and Bardin 
{\it et al.}, respectively.}
\end{Abstract}
\vfill
\begin{Presented}
{\Color{Orange}5th International Symposium on Radiative Corrections \\ 
(RADCOR--2000) \\[4pt]
Carmel CA, USA, 11--15 September, 2000}
\end{Presented}
\vfill
\end{titlepage}
\def\thefootnote{\arabic{footnote}}
\setcounter{footnote}{0}

\section{Introduction}
%%%Put a different spin on this paragraph.
The theoretical paradigm affirmed by the award of the 1999 Nobel Prize to
G. 't Hooft and M. Veltman for the success of the predictions
of their formulation~\cite{gthmv} of the renormalised non-Abelian quantum
loop corrections for the {\Color{Red}Standard Model~\cite{gsw}} 
of the electroweak
interaction focuses our efforts on the need to continue to test 
this theory at the quantum loop level in the gauge boson sector
itself. This then emphasises the importance of the on-going
(the data are under analysis and will be for some time
even though the LEP2 accelerator was recently shutdown) 
precision studies of the processes
%%WP $e^+e^- \to W^+W^- +n(\gamma)\to 4 fermions+n(\gamma)$
{\Color{Red}$e^+e^- \to W^+W^-(ZZ) +n(\gamma)\to 4f+n(\gamma)$}
at LEP2
energies~\cite{lep2ybk:1996,frits:1998,frits:1999}, 
as well as the importance of the planned future
higher energy studies of such processes in
LC physics 
programs~\cite{nlc:1995,jlc:1995,tesla:1998,tesla:2000}.
We need to stress also that hadron colliders also have considerable
reach into this physics and we hope to come back to their
roles elsewhere~\cite{kkcol1}.\par
In what follows, we present precision predictions for the
event selections (ES) of the LEP2 MC Workshop~\cite{lep2YR:2000}
for the processes {\Color{Green}$e^+e^- \to W^+W^- +n(\gamma)\to 4f+n(\gamma)$}
based on our new exact {\Color{Blue}${\cal O}(\alpha)_{prod}$} 
YFS exponentiated
{\Color{Blue}LL ${\cal O}(\alpha^2)$ FSR} 
{\Color{Maroon}leading pole approximation (LPA)} formulation
as it is realized in the MC program 
{\Color{Red}YFSWW3-1.14 ~\cite{yfsww3:1998,yfsww3:2000}} in 
combination with the all four-fermion processes 
MC event generator {\Color{Red}KoralW-1.42~\cite{krlw:1999}}
so that the respective four-fermion background processes are
taken into account in a gauge invariant way. 
In addition, we also present the current status of the 
predictions of our YFS MC approach to the processes
{\Color{Red}$e^+e^- \to ZZ +n(\gamma)\to 4f+n(\gamma)$} 
as it was also illustrated
in the {\Color{Orange}2000 LEP2 MC Workshop~\cite{lep2YR:2000}} using the
MC event generator YFSZZ~\cite{yfszz:1997}, which realizes
{\Color{Magenta}YFS exponentiated} {\Color{Red}LL ${\cal O}(\alpha^2)$ 
ISR in the LPA}
in a gauge invariant way. Indeed, gauge invariance
is a crucial aspect of our work and we stress that we maintain
it through-out our calculations. Here, {\Color{Red}ISR} denotes 
initial state radiation, {\Color{Red}FSR} denotes final-state
radiation and LL denotes leading-log as usual.\par

This realization which we present of the YFS MC approach
is the {\Color{Red}exclusive exponentiation (EEX)~\cite{yfs:1961}} and it is
already well established in its applications to the
{\Color{Magenta}MC event generators for LEP1 physics calculations
in the MC's KORALZ/YFS3~\cite{koralz4:1994,yfs3:1992}, 
BHLUMI~\cite{bhlumi:1994,bhlumi:1992} and KoralW~\cite{krlw:1999}}.
In our applications in YFSWW3-1.14 and in KoralW-1.42, the {\Color{Blue}FSR
is implemented using the program PHOTOS~\cite{photos:1994}}, so that
not only is the FSR calculated to the {\Color{Green}LL ${\cal O}(\alpha^2)$
but the FSR photons have the correct finite $p_T$ in the soft
limit to ${\cal O}(\alpha)$}. We always use the ratio of
branching ratios (BR's) to correct the respective
decay rates through {\Color{Blue}${\cal O}(\alpha)$} accordingly.
Recently, we have introduced the {\Color{Maroon}coherent exclusive 
exponentiation (CEEX)~\cite{ceex}} approach
to the YFS MC event generator calculation of radiative corrections
and we will present the application of this new approach to the $4f$ production
processes elsewhere~\cite{kkcol1}. {\Color{Blue}For a description
of its application to the $2f$ production processes see Ref.~\cite{zwas}}.\par

Recently, the authors in Refs.~\cite{racnw:1999,dittm} have also 
presented MC 
program
results for the processes
{\Color{Red}$e^+e^- \to W^+W^- +n(\gamma)\to 4f+n(\gamma),~n=0,1$}
in combination with the complete background processes
which feature the {\Color{Green}exact LPA  ${\cal O}(\alpha)$ correction,
the complete ${\cal O}(\alpha)$ result for 
$e^+e^-\to 4f+\gamma$, and soft photon KF~\cite{kf-expn}
exponentiation for the LL ${\cal O}(\alpha^3)$ ISR via
structure functions}.
Thus, we will compare our results where possible with those in
Refs.~\cite{racnw:1999} in an effort to check the over-all
precision of our work. As we argue below, the two sets of results
should agree at a level below $0.5\%$ on observables such as the
total cross section. The authors in Refs.~\cite{frits:1999}
have used {\Color{Magenta}semi-analytical methods} to compute the 
{\Color{Maroon}exact LPA  
${\cal O}(\alpha)$ correction $e^+e^- \to W^+W^- \to 4f$}
{\Color{Green}with no higher order resummation}. Thus, while we have compared
our results with theirs in Ref.~\cite{yfsww3:2000} for example,
here we do not present such comparisons because {\Color{Red}the expected
precision tag of their results is larger than the desired
$0.5\%$ needed by the LEP2 experiments~\cite{lep2YR:2000}}.
\par

For the processes {\Color{Red}$e^+e^-\to ZZ+(n\gamma)\to 4f+(n'\gamma)$},
the authors in Refs.~\cite{pass:2000,bard:2000} have presented
calculations in the LEP2 MC Workshop~\cite{lep2YR:2000}
at the {\Color{Magenta}NC02 and all-$4f$ level~\cite{bard:2000}}
as well. The calculations {\Color{Magenta}in Ref.~\cite{pass:2000}
are done with the program ZZTO and feature universal ISR corrections,
${\cal O}(\alpha)$ FSR$_{QED}$ corrections, 
${\cal O}(\alpha_s)$ FSR$_{QCD}$ corrections, and running masses
in the fermion loop scheme of Ref.~\cite{fl-scheme}}.
The results {\Color{Magenta}in Ref.~\cite{bard:2000} feature the structure
function approach to the ISR QED corrections and the
${\cal O}(\alpha)$ FSR$_{QED}$ corrections}. We will compare our
YFSZZ results with these two sets of results as well, as the
three approaches should agree {\Color{Red}at the level of the 2\% precision
needed by the LEP2 experiments~\cite{lep2YR:2000}
on observables such as the total cross section}.
\par

Our presentation is organised as follows. {\Color{Red}In the next Section},
we discuss the current status YFSWW3-1.14. {\Color{Red}In Section 3},
we present the current status of KoralW-1.42 from the standpoint of
its use to calculate the $4f$ background processes in combination
with YFSWW3-1.14. {\Color{Red}In Section 4}, we present the current status
of YFSZZ. {\Color{Red}In Sections 5 , 6 and 7}, 
we illustrate the results we have obtained
with our calculations for YFSWW3, KoralW-1.42 and YFSZZ, respectively,
for the ES of the LEP2 MC Workshop~\cite{lep2YR:2000},
wherein we include {\Color{Maroon}comparisons with the respective results 
in Refs.~\cite{racnw:1999,pass:2000,bard:2000}}. {\Color{Red}Section 8}
contains our summary remarks.

\section{YFSWW3-1.14}

In this section we present the current status of 
{\Color{Maroon} YFSWW3-{\Color{PineGreen}1.14}}. We start with the
{\bf\Color{Blue}
{\Color{Red} process of interest} and its cross section,
\begin{equation}
  \label{4fprocess}
  \begin{split}
    &{\Color{Black}e^-(p_1) + e^+(p_2) 
    \to  f_1(r_1)+\bar{f}_2(r_2) + f'_1(r'_1)+\bar{f}'_2(r'_2)} 
    + {\Color{Orange}\gamma(k_1),...,\gamma(k_n)},\\
   &\sigma_n = {1\over {\Color{Black}flux}}
    \int d\tau_{n+4}({\Color{Black}p_1+p_2;r_1,r_2,r'_1,r'_2},{\Color{Orange}k_1,...,k_n})\\
   &\qquad\qquad
    \sum_{{\Color{Black}ferm.~spin}}\; \sum_{{\Color{Orange}phot.~spin}} 
      |{\cal M}^{(n)}_{{\Color{Red}4f}}({\Color{Black}p_1,p_2,r_1,r_2,r'_1,r'_2},{\Color{Orange}k_1,...,k_n})|^2,
  \end{split}
\end{equation}
and the corresponding expressions for the 
{\Color{Red}$W^+W^-$ production and decay} in the leading pole 
approximation {\Color{Maroon}(LPA)},
\begin{equation}
  \label{WWprod-decay}
  \begin{split}
   &{\Color{Black}e^-(p_1) + e^+(p_2)} \to  {\Color{Red}W^-(q_1) + W^+(q_2)},\quad \\
   &{\Color{Red}W^-(q_1)} \to {\Color{Black}f_1(r_1)+\bar{f}_2(r_2)},\quad 
    {\Color{Red}W^+(q_2)} \to {\Color{Black}f'_1(r'_1)+\bar{f}'_2(r'_2)},\\
   &\sigma_n = {1\over {\Color{Black}flux}}
    \int d\tau_{n+4}({\Color{Black}p_1+p_2;r_1,r_2,r'_1,r'_2},{\Color{Orange}k_1,...,k_n})\\
   &\qquad\qquad
    \sum_{{\Color{Black}ferm.~spin}}\; \sum_{{\Color{Orange}phot.~spin}}
      |{\cal M}^{(n)}_{{\Color{Red}LPA}}({\Color{Black}p_1,p_2,r_1,r_2,r'_1,r'_2},{\Color{Orange}k_1,...,k_n})|^2.
  \end{split}
\end{equation}
Here, we realize the 
{\Color{Maroon} LPA$_{\Color{Magenta}a,b}$ as follows:}
{\small
\begin{equation}
  \label{eq:lpa}
  \begin{split}
   &{\cal M}^{(n)}_{{\Color{Red}4f}}({\Color{Black}p_1,p_2,r_1,r_2,r'_1,r'_2},{\Color{Orange}k_1,...,k_n})\;\;
    {{\Color{Maroon}LPA} \atop => }
   {\cal M}^{(n)}_{{\Color{Red}LPA}}({\Color{Black}p_1,p_2,r_1,r_2,r'_1,r'_2},{\Color{Orange}k_1,...,k_n}) \\
   &=\sum_{{\Color{Orange}Phot.~Partitions}}\;
   {\cal M}^{(n),{\Color{Red}\lambda_1 \lambda_2}}_{{\Color{Maroon}Prod}}({\Color{Black}p_1,p_2},{\Color{Red}q_1,q_2},{\Color{Orange}k_1,...,k_a})\\
   &\times {1\over {\Color{Red}D(q_1)}}\; 
       {\cal M}^{(n)}_{{\Color{Maroon}Dec_1},{\Color{Red}\lambda_1}}({\Color{Red}q_1},{\Color{Black}r_1,r_2},{\Color{Orange}k_{a+1},...,k_b})\; \\
    &\times {1\over {\Color{Red}D(q_2)}}\; 
       {\cal M}^{(n)}_{{\Color{Maroon}Dec_2},{\Color{Red}\lambda_2}}({\Color{Red}q_2},{\Color{Black}r'_1,r'_2},{\Color{Orange}k_{b+1},...,k_n}),\\
   &{\Color{Red}D(q_i) = q_i^2-M^2},\quad
    M^2 = (M_{\Color{Red}W}^2-i\Gamma_{\Color{Red}W} M_{\Color{Red}W})(1-\Gamma_{\Color{Red}W}^2/M_{\Color{Red}W}^2 +{\cal O}({\Color{Magenta}\alpha^3})),\\
   &{\Color{Red}q_1}= {\Color{Black}r_1+r_2}+{\Color{Orange}k_{a+1}+...+k_b};\;\; {\Color{Red}q_2}={\Color{Black}r'_1+r'_2}+{\Color{Orange}k_{b+1}+...+k_n},
  \end{split}
\end{equation}}
where the two formulations of the {\Color{Maroon} LPA},
{\Color{Maroon} LPA$_{{\Color{Magenta}a},{\Color{Black}b}}$},
are based on the results in 
{\Color{Green} Eden Refs.~\cite{eolp:1966,strt:1995}}
%%Eden {\it et al.}, Stuart, hep-ph/9706431, etc.}
as one can see from the representation of our amplitudes ${\cal M}$ as
\begin{equation}
{\cal M} = \sum_j {\Color{Black}\ell_j}{\Color{Magenta}A_j\left(\{q_kq_l\}\right)}.
\end{equation} 
Here, the $\{\ell_j\}$ are a complete set of spinor covariants
and the $\{A_j\}$ are the respective scalar functions. For 
{\Color{Maroon}LPA$_{({\Color{Magenta}a}){\Color{Black}b}}$, we do (not) 
evaluate the
spinor covariants on-pole in realizing the respective ${\cal M}^{(n)}_{LPA}$.
{\Color{Red} We do both in {\Color{Maroon}YFSWW3-1.14}.}}
\par
We use {\Color{Black}standard YFS methods{\Color{Red}(EEX-Type)}}{\Color{PineGreen}to write}
\begin{equation}
  \label{eq:archaic2}
  \begin{split}
   &d\sigma =
    e^{2\Re {\Color{Magenta}\alpha} {\Color{Orange}B'} +2{\Color{Magenta}\alpha} {\Color{Orange}\tilde{B}}} {1\over (2\pi)^4}\\
   &\int d^4 y e^{iy({\Color{Black}p_1+p_2}{\Color{Red}-q_1-q_2})+{\Color{Orange}D}}
    [\bar{\beta}_{\Color{Orange}0} +\sum_{{\Color{Orange}n=1}}^{\Color{Orange}\infty} {d^3 {\Color{Orange}k_j}\over {\Color{Orange}k^0_j}} e^{-iy{\Color{Orange}k_j}}
    \bar{\beta}_{\Color{Orange}n}({\Color{Orange}k_1,...,k_n})]\\
&\qquad \times{d^3{\Color{Black}r_1} \over {\Color{Black}\bar E_1}}{d^3{\Color{Black}r_2} \over {\Color{Black}\bar E_2}}
  {d^3{\Color{Black}r'_1} \over {\Color{Black}\bar E'_1}} {d^3{\Color{Black}r'_2} \over {\Color{Black}\bar E'_2}},
\end{split}
\end{equation}
{\Color{Red}where}
\begin{equation}
\label{eq:ryfsfn}
 \begin{split}
 {\Color{Orange}D}&=\int{d^3{\Color{Orange}k}\over
{\Color{Orange}k_0}}\tilde S\left[e^{-iy\cdot {\Color{Orange}k}}-\theta({\Color{Orange}K_{max}}-|{\Color{Orange}\vec k}|)\right]\\
 2{\Color{Magenta}\alpha} {\Color{Orange}\tilde B} &=\int{d^3{\Color{Orange}k}\over {\Color{Orange}k_0}}\theta({\Color{Orange}K_{max}}-|{\Color{Orange}\vec k}|)\tilde S({\Color{Orange}k}). 
\end{split}
\end{equation}
Here,
{\Color{Red} $K_{max}$ is a dummy parameter of which eq.(\ref{eq:archaic2})
is independent}.
In realizing eq.(\ref{eq:archaic2}) in YFSWW3, we employ the following 
{\Color{Maroon}schemes, which are related by the renormalization group:}
\begin{itemize}
\item
Version 1.13: {\Color{Magenta}$G_\mu$-Scheme} of 
Fleischer {\it et al.}~\cite{ew1}
% {\Color{Magenta}Z. Phys. {\bf C42} (1989) 409, etc.}
\item
Version 1.14: {\Color{PineGreen}Scheme A} -- only the hard EW correction
has {\Color{Magenta}$\alpha_{G_\mu}$};
{\Color{Orange}Scheme B} -- the entire {\Color{Magenta}${\cal O}(\alpha)$} 
correction has {\Color{Magenta}$\alpha(0)$}
\end{itemize}
As it was shown in the LEP2 MC Workshop~\cite{lep2YR:2000},
there is a 
{\Color{Orange} $\Rightarrow$} {\Color{Magenta}$-0.3\div-0.4\%$} 
shift of the {\Color{Red}normalisation} of {\Color{Black}version 1.14}
relative to that of {\Color{Black}version 1.13}.
This can be seen as follows.
The universal LL ISR ${\cal O}(\alpha)$ soft plus virtual correction is
%{\Color{Green} See Dittmaier's Talk for More Details and References}
}
\begin{equation}
\label{eq:s+v}
\delta^{v+s}_{ISR,LL}{\Color{Red}=}{\Color{Magenta}\beta}\ln {\Color{Orange}k_0}+\frac{\alpha}{\pi}\left(\frac{3}{2}{\Color{Magenta}L}+\frac{\pi^2}{3}-2\right),
\end{equation}
{\Color{Green}with}
${\Color{Magenta}\beta}{\Color{Red}=}\frac{2\alpha}{\pi}({\Color{Magenta}L}-1)$
and with $k_0$ equal to the usual soft cut-off and $L=\ln s/m_e^2$.
From eq.(\ref{eq:s+v}), we get the estimate of the 
{\Color{Brown}shift in normalisation
between version 1.13 and version 1.14 at 200 GeV} as
\begin{equation} 
(\alpha(0)-\alpha_{G_\mu})(\frac{3}{2}{\Color{Magenta}L}-2){\Color{Red}\sim -0.33\%}.
\end{equation} 
This is consistent with what is observed as reported 
in Ref.~\cite{lep2YR:2000}. 
{\Color{Green}See Dittmaier's talk~\cite{dittm} for 
more details and references.}
\par
\section{KoralW-1.42}

For the
{\bf\Color{Blue}
{\Color{Red} process of interest},
${\Color{Black}e^-(p_1) + e^+(p_2) 
    \to  f_1(r_1)+\bar{f}_2(r_2) + f'_1(r'_1)+\bar{f}'_2(r'_2)} 
    + {\Color{Orange}\gamma(k_1),...,\gamma(k_n)}$,
{\Color{Red}we
use KoralW-1.42 which realizes the ${\cal O}(\alpha^3)$ LL YFS exponentiated
ISR}. The respective input Born matrix elements are the GRACE v. 2~\cite{jfuj}
%{\Color{Black}(J. Fujimoto {\it et al.},
%MINAMI-TATEYA Coll., GRACE User's Manual, v. 2.0)} 
all 4f library of Born matrix elements and {\Color{Green} our independent CC03 Born matrix elements}.
This allows us to {\Color{Red}combine YFSWW3-1.14 and KoralW-1.42 to correct 
for background diagram effects}: using LPA$_a$ in YFSWW3-1.14, whose
cross section we denote by
{\Color{Magenta}$\sigma(Y_a)$}, we get
\begin{equation}
\sigma_{Y/K}= {\Color{Magenta}\sigma(Y_a)}+\Delta\sigma({\Color{Magenta}K}),
\label{sigya}
\end{equation}
where $\Delta\sigma({\Color{Magenta}K})$ is defined by 
\begin{equation}
\Delta\sigma({\Color{Magenta}K})= \sigma({\Color{Magenta}K}_1)-\sigma({\Color{Magenta}K}_3).
\end{equation}
Here,  $\sigma({\Color{Magenta}K}_1)$  {\Color{Red}is the} {\Color{Black}4-f} KoralW-1.42 result and
 $\sigma({\Color{Magenta}K}_3)$  {\Color{Red}is the} {\Color{Black}CC03} KoralW-1.42 result.
{\Color{Black}This means that} ${\Color{Red}\sigma_{Y/K}}$ is accurate to 
{\Color{Red}${\cal O}(\frac{\alpha}{\pi}\frac{\Gamma_W}{M_W})$}.\par

Alternatively, using  LPA$_i,~i=a,b$ in YFSWW3-1.14,
whose cross section we denote by {\Color{Magenta}$\sigma(Y_i)$}, we get
\begin{equation}
\sigma_{K/Y} = {\Color{Magenta}\sigma(K_1)}+\Delta\sigma({\Color{Magenta}Y})
\end{equation}
where
\begin{equation}
\Delta\sigma({\Color{Magenta}Y})=\sigma({\Color{Magenta}Y}_i)-\sigma({\Color{Magenta}Y}_4),
\end{equation}
and 
$\sigma({\Color{Magenta}Y}_4)$ {\Color{Black}is the respective}
YFSWW3-1.14 result with
{\Color{Magenta}NL ${\cal O}(\alpha)$ corrections to $\bar\beta_n$, $n=0,1$, switched off}.
{\Color{Black}This means that} ${\Color{Red}\sigma_{K/Y}}$ is also accurate to
{\Color{Red}${\cal O}(\frac{\alpha}{\pi}\frac{\Gamma_W}{M_W})$}.\par
{\Color{Green}
Above WW threshold, {\Color{Magenta}$\sigma_{K/Y}$} and  {\Color{Magenta}$\sigma_{Y/K}$}\\
agree to the {\Color{Red}$0.1\%$ level}.
We advocate the latter as our best 
result in the following.}\par
Note that we sometimes identify
{\Color{Orange}
$\sigma(Y_1)=\sigma(Y_a),~\sigma(Y_2)=\sigma(Y_b),~\sigma(Y_3)=\sigma(K_3)$}
with
{\Color{Orange}$\sigma(K_2)$} equal to the cross section from KoralW-1.42
with the {\Color{Orange}on-pole} CC03 Born level matrix element
with {\Color{Red}YFS exponentiated ${\cal O}(\alpha^3)$ LL ISR}
-- this {\Color{Orange}$\sigma(K_2)$} should be available soon.
It is useful for further cross checks on our work.
}\par
\section{YFSZZ}

In our calculation in {\Color{Maroon} 
YFSZZ-{\Color{PineGreen}1.02~\cite{yfszz:1997}}}
{\bf\Color{Blue}the 
{\Color{Red} process of interest} is
${\Color{Black}e^-(p_1) + e^+(p_2)}$ {\Color{Blue}$\to$} 
${\Color{Black}Z(q_1) Z(q_2)+({\Color{Orange}\gamma(k_1),...,\gamma(k_m)})}$
${\Color{Black} {\Color{Blue}\to}  f_1(r_1) + \bar{f}_1(r_2) + f'_1(r'_1) + \bar{f}'_1(r'_2)} 
 + {\Color{Orange}\gamma(k_1),...,\gamma(k_n)}$.
We proceed as follows in realizing the MC {\Color{Maroon} 
YFSZZ-{\Color{PineGreen}1.02}}:
\begin{itemize}
\item
We use {\Color{PineGreen}$LPA_a$} as described above for the NC02 process
to calculate
{\Color{Red}${\cal O}(\alpha^2)$ LL YFS exponentiated
ISR} for the input {\Color{Green} 
%Hagiwara, 
%{\it et al.}(NPB{\bf 282}, 253 (1987)) 
NCO2 Born matrix elements
of Ref.~\cite{hagi}.}
\item
{\Color{Black} Anomalous couplings are supported following 
the conventions of Ref.~\cite{hagi}}
-- this is also true for {\Color{Orange}YFSWW3/KORALW}.
\end{itemize}
\par
We stress that YFSZZ is in wide use at LEP and that it
{\Color{Red} was tested in the LEP2 MC Workshop}, just as YFSWW3-1.14 
was tested.
{\Color{Magenta}We now turn to such results.}
}
\section{Results-YFSWW3-1.14}

\begin{figure}[b!]
\begin{center}
{\Color{Blue}    Hardest Photon Angular Distribution}

%\begin{center}
 
%\framebox{\Large{\Color{Black}
% $\: e^+e^- \longrightarrow W^+W^- \longrightarrow
%{\Color{Red} u \bar{d}} \mu^- \bar{\nu}_{\mu}$}
%}
 
\setlength{\unitlength}{0.0625mm}
\begin{picture}(1600, 780)
%%%%%%%%%%%%%%%%%%%%%%%%%%%%%%%%%%
\put( 450, 720){\makebox(0,0)[cb]{$E_{CM} = 200\,GeV$} }
\put(1250, 720){\makebox(0,0)[cb]{$E_{CM} = 500\,GeV$} }

\put(   0,  0){\makebox(0,0)[lb]{
\epsfig{file=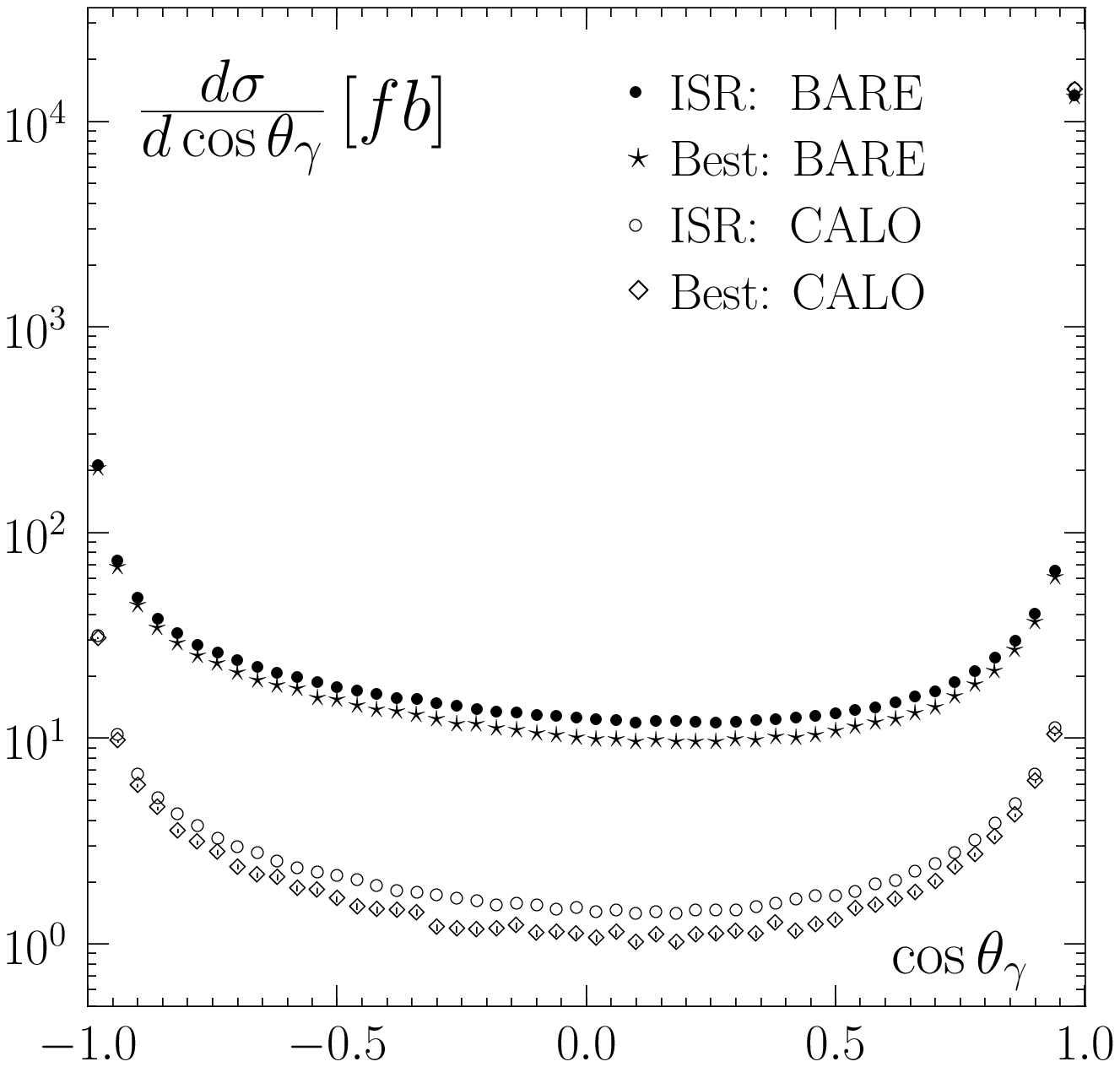       ,width=50mm,height=43.75mm}
}}
%#####

\put( 800,  0){\makebox(0,0)[lb]{
\epsfig{file=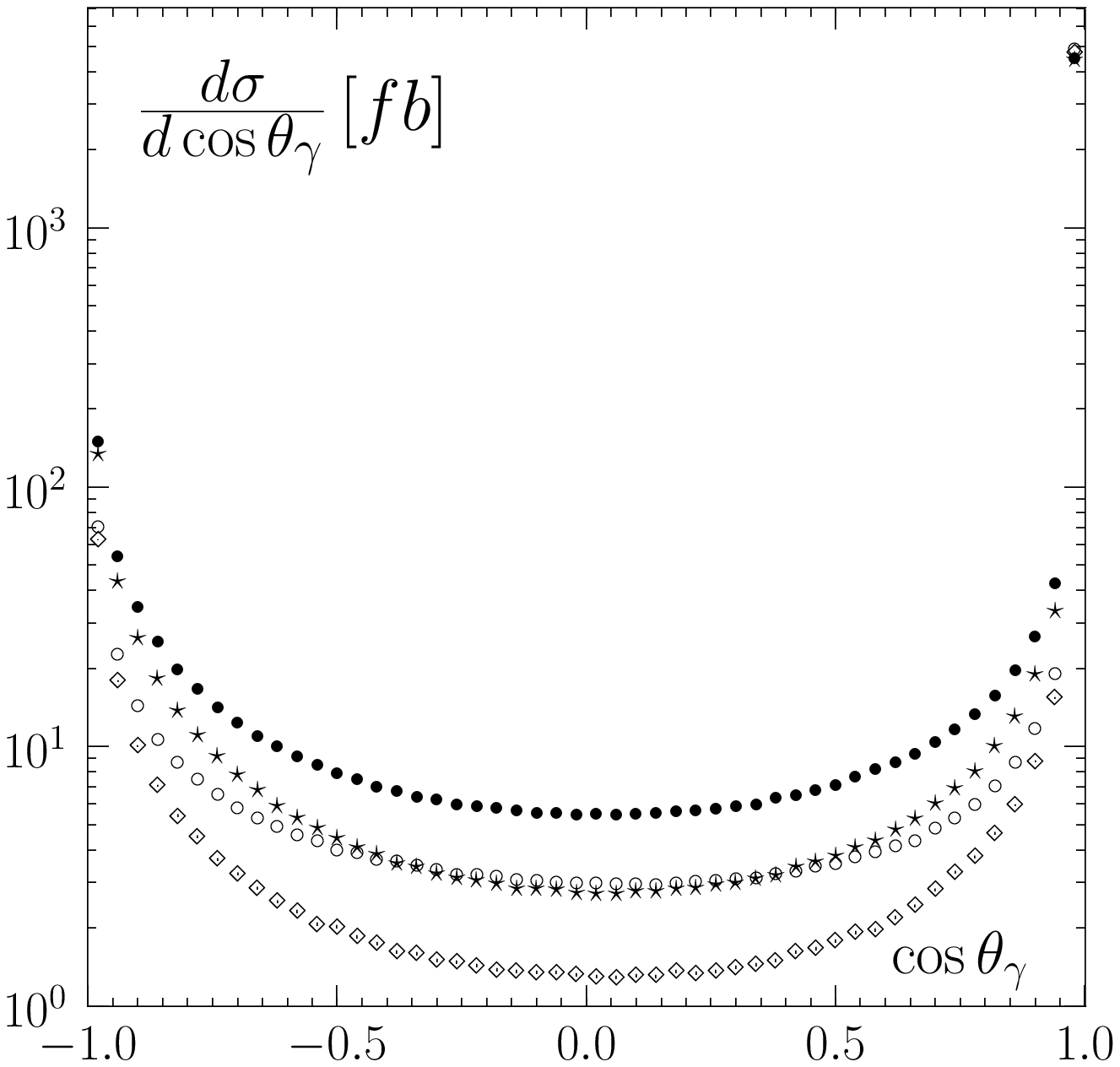       ,width=50mm,height=43.75mm}
}}

%
%\setlength{\unitlength}{1mm}
%\begin{picture}(100,60)
%%%\put(0,0){\framebox( 65,60){ }}
%\put(0, 00){\makebox(0,0)[lb]{
%\epsfig{file=chi-mcan-O0tech.999.eps,width=80mm,height=62mm}
%}}
\end{picture}
\end{center}
\caption[0]{\Color{Orange}$cos\theta_\gamma$ {\Color{Black}w.r.t. the $e^+$ beam in the cms system for {\Color{Black}
 $e^+e^- \longrightarrow W^+W^- \longrightarrow
{\Color{Red} u \bar{d}} \mu^- \bar{\nu}_{\mu}$}.
} {\Color{PineGreen}We see that} {\Color{Blue}NL} corrections 
are important {\Color{Red}away} from the beams, for example.}
\label{fig:cgcmsd}
\end{figure}

{\Color{Blue}In this section we illustrate the effects of the NL
${\cal O}(\alpha)$ correction as it is calculated in YFSWW3-1.14.
We do this with the hardest photon angular distribution.
Similar calculations of other observables can be found 
in Ref.~\cite{yfsww3:2000,lep2YR:2000}.\par

Specifically, in Fig.~\ref{fig:cgcmsd}, we show the 
distribution of the cosine of the production angle of the
hardest photon in the cms system with respect to the $e^+$ beam.
We see that away from the beams the NL ${\cal O}(\alpha)$ correction
is important for precision studies of this photonic observable.
Similar conclusions follow from the more complete set of
observables studied in Refs.~\cite{yfsww3:2000,lep2YR:2000}.}
%//////////////////////////////////////////////////////////////////////////////////
%\begin{slide}\titbox{{\large\bf\Color{Red}RESULTS: YFSWW3-1.14}}
\par
{\Color{Maroon}Indeed, in the LEP2 MC Workshop, we compared our results
with those of RacoonWW by the authors in Ref.~\cite{racnw:1999}.
For a complete description of these comparisons we refer
the reader to Ref.~\cite{lep2YR:2000}. Here, we show
in Table~\ref{ta:totcsYFSRacnocuts} the comparison
of the total cross sections at 200 GeV with no cuts as defined
Ref.~\cite{lep2YR:2000}.} 
{\Color{Blue}  
%\begin{center}
%\end{center}
\let\sstl=\scriptscriptstyle
%%%=================local=macros=========================
\def\Was{W\c as}
\def\Order#1{${\cal O}(#1$)}
\def\Ordpr#1{${\cal O}(#1)_{prag}$}
\def\bbe{\bar{\beta}}
\def\tbe{\tilde{\beta}}
\def\tal{\tilde{\alpha}}
\def\tom{\tilde{\omega}}
\def\half{ {1\over 2} }
\def\alf1{ {\alpha\over\pi} }

\def\Oaz{${\cal O}(\alpha^0)$}
\def\Oaf{${\cal O}(\alpha^1)$}
\def\Oas{${\cal O}(\alpha^2)$}
%%%======================================================
% ============ begin table 1 ===============
{\small
\begin{table}[t]
\begin{center}
{\Color{Orange} Comparison with} RacoonWW\nl
\begin{tabular}{|c|c|c|c|}
\hline
\multicolumn{2}{|c|}{\bf {\Color{Red}no cuts}}&
\multicolumn{2}{|c|}{\bf{\Color{Black}$\sigma_{\mathrm{tot}}[\mathrm{fb}]$}}\nl
\hline
final state & program & {\Color{Maroon}Born} & {\Color{PineGreen}best} \nl
\hline\hline
& {\tt {\Color{Magenta}YFSWW3}} & {\Color{Magenta}219.770(23)} & {\Color{Magenta}199.995(62)} \nl
$\nu_\mu\mu^+\tau^-\bar\nu_\tau$
& {\tt RacoonWW} & 219.836(40) & 199.551(46) \nl
\cline{2-4}
& ({\Color{Magenta}Y}--R)/{\Color{Magenta}Y} & {\Color{Maroon}$-0.03(2)$\%} &  {\Color{PineGreen}0.22(4)\%} \nl
\hline\hline
& {\tt {\Color{Magenta}YFSWW3}} & {\Color{Magenta}659.64(07)} & {\Color{Magenta}622.71(19)} \nl
$\Pu\bar\Pd\mu^-\bar\nu_\mu$
& {\tt RacoonWW} & 659.51(12) & 621.06(14) \nl
\cline{2-4}
& ({\Color{Magenta}Y}--R)/{\Color{Magenta}Y} & {\Color{Maroon}$0.02(2)$\%} &  {\Color{PineGreen}0.27(4)\%} \nl
\hline\hline
& {\tt {\Color{Magenta}YFSWW3}} & {\Color{Magenta}1978.18(21)} & {\Color{Magenta}1937.40(61)} \nl
$\Pu\bar\Pd\Ps\bar\Pc$
& {\tt RacoonWW} & 1978.53(36) & 1932.20(44) \nl
\cline{2-4}
& ({\Color{Magenta}Y}--R)/{\Color{Magenta}Y} & {\Color{Maroon}$-0.02(2)$\%} &  {\Color{PineGreen}0.27(4)\%} \nl
\hline
\end{tabular}
\end{center}
%\end{table}
%}
\caption{
%\end{center}
{\Color{Orange}Total cross sections, CC03 from {\tt {\Color{Blue}RacoonWW}},
{\tt  {\Color{Magenta}YFSWW3}}, $\sqrt{s}=200\,\GeV$ 
{\Color{Red}without cuts}. Statistical errors 
correspond to the last digits in $(~)$.}}
\label{ta:totcsYFSRacnocuts}
\end{table}
}
%\begin{center}
%%\setlength{\unitlength}{1mm}
%%\begin{picture}(100,60)
%%%%\put(0,0){\framebox( 65,60){ }}
%%\put(0, 00){\makebox(0,0)[lb]{
%%\epsfig{file=chi-mcan-O0tech.999.eps,width=80mm,height=62mm}
%%}}
%-----------------------------------------------------------------------------------------
%\begin{figure}[!ht]
%%\centering
%\end{center}
%\end{picture}
}
From the results in Table~\ref{ta:totcsYFSRacnocuts} 
and the related results given in Ref.~\cite{lep2YR:2000}
we conclude that the TU of the calculations is {\Color{Blue}0.4\% at 200 GeV}
for the total signal cross section. 
{\Color{Red}This is a considerable improvement
over previously quoted precision of ~2\% in Ref.~\cite{previous}.}\par 
\section{Results-YFSWW3/KoralW}
{\Color{Blue}One of the important aspects of the isolation and study of the
$WW$ signal processes is the control of the corresponding 
{\Color{Red}background
$4f$ processes}. This we do with our all-$4f$ MC KoralW-1.42 as
we described above. Here, we illustrate the size of the corresponding
$4f$ background corrections to the YFSWW3-1.14 cross sections.}\par

Specifically, in Tabs.~\ref{YK-nocuts} and \ref{YK-cuts},
we show the size of this {\Color{Red}$4f$ background correction in comparison
to the NL correction of YFSWW3-1.14 for the total cross section}, 
for example, both for
the case of no cuts and the case of cuts, respectively,
as defined in Ref.~\cite{lep2YR:2000}. {\Color{Red}These results show that the
$4f$ background correction at 200 GeV to the total YFSWW3-1.14 cross section
is below 0.1\%.}
{\Color{Blue}
{\small
\begin{table}[t]
\begin{center}
{\Color{Orange}$WW/4f$ Cross Section}\nl
%\begin{tabular}{|c|c|c|c|}
%\begin{table*}[hbtp]
%\centering
\begin{tabular}{||c|c||c|c||c|c||c||}
\hline\hline
\multicolumn{2}{||c||}{\bf {\Color{Red}NO CUTS}}  & 
\multicolumn{2}{c||}{$\sigma_{{\Color{Red}WW}}\,[fb]$} &
\multicolumn{2}{c||}{$\delta_{{\Color{Magenta}4f}}\,[\%]$} &
\raisebox{-1.5ex}[0cm][0cm]{$\delta_{{\Color{Red}WW}}^{NL}\,[\%]$} \\
%\hline
\cline{1-6}{\Color{Orange}
Final state} & {\Color{Orange}Program} & {\Color{Orange}Born} & {\Color{Orange}ISR} & {\Color{Orange}Born} &  {\Color{Orange}ISR} & \\
\hline\hline
& 
{\Color{Black}YFSWW3} & 
$219.793 \,(16)$ &
$204.198 \,(09)$ &
--- & --- &
$-1.92 \,(4)$ \\
{\Color{Green}$\nu_{\mu}\mu^+\tau^-\bar{\nu}_{\tau}$} &
{\Color{Maroon}KoralW} & 
$219.766\,(26)$ &
$204.178\,(21) $ &
${\Color{Magenta}0.041} $ &
${\Color{Magenta}0.044} $ & 
--- \\
\cline{2-4}
&
(Y$-$K)/Y &
$0.01 \,(1)\% $ &
$0.01 \,(1)\% $ &
--- & --- & --- \\
\hline\hline
 & 
{\Color{Black}YFSWW3} & 
$659.69 \,(5)$ &
$635.81 \,(3)$ &
--- & --- &
$-1.99 \, (4)$ \\
{\Color{PineGreen}$u\bar{d}\mu^-\bar{\nu}_{\mu}$} &
{\Color{Maroon}KoralW} & 
$659.59 \,(8)$ &
$635.69 \,(7)$ &
$ {\Color{Magenta}0.073} $ &
$ {\Color{Magenta}0.073} $ & 
--- \\
\cline{2-4}
&
(Y$-$K)/Y &
$ 0.02 \,(1)\% $ &
$ 0.02 \,(1)\% $ &
--- & --- & --- \\
\hline\hline
 & 
{\Color{Black}YFSWW3} & 
$1978.37 \, (14)$ &
$1978.00 \, (09)$ &
--- & --- &
$-2.06 \,(4)$ \\
{\Color{Brown}$u\bar{d} s\bar{c} $} &
{\Color{Maroon}KoralW} & 
$1977.89 \, (25) $ &
$1977.64 \, (21) $ &
$ {\Color{Magenta}0.060} $ &
$ {\Color{Magenta}0.061} $ & 
--- \\
\cline{2-4}
&
(Y$-$K)/Y  &
$ 0.02 \,(1)\% $ &
$ 0.02 \,(1)\% $ &
--- & --- & --- \\
\hline\hline
\end{tabular}
%\caption{\sf
%The total $WW$ cross sections from YFSWW3 and KoralW 
%at the Born and ISR level, the $4f$ corrections from KoralW and
%${\cal O}(\alpha)$ NL correction from YFSWW3 at $E_{CM} = 200\,GeV$, 
%without cuts. 
%The numbers in parentheses are statistical
%errors of the results corresponding to last digits.
%}
%\label{tab:YR-xtot-nocuts}
%\end{table*}
%%%%%%%%%%%%%%%%%%%%%%%%%%%%%%%%%%%%%%%%%%%%%%%%%%%%%%%%%%%%%%%%%%%%%%%%%%%%
\end{center}
%\end{table}
%}
\caption{
%\end{center}
\Color{Orange}Total ${\Color{Red}WW}$  {\Color{Black}YFSWW3} and {\Color{Maroon}KoralW} cross sections: 
Born and ISR level, 
{\Color{Maroon}KoralW} ${\Color{Magenta}4f}$ correction,
{\Color{Black}YFSWW3} ${\cal O}(\alpha)$ NL correction, at $200\,GeV$, 
no cuts. The 
{\Color{Red}last digits in $(\cdots)$} correspond to the statistical errors.}
\label{YK-nocuts}
\end{table}
}
%\begin{figure}[b!]
%\begin{center} 
%%{\Color{Orange}    WW/4f Cross Section}\nl
%%
%\setlength{\unitlength}{1mm}
%\begin{picture}(120,80)
%\put(45,50){\Color{Orange}$WW/4f$ Cross Section}
%%%\put(0,0){\framebox( 65,60){ }}
%\put(10,-40){\makebox(10,-40)[lb]{
%\epsfig{file=tabyvk-xtot.eps,width=100mm,height=82mm}
%}}
%\end{picture}
%\end{center}
%\vskip-.01cm
%\caption{\small Total ${\Color{Red}WW}$  {\Color{Black}YFSWW3} and {\Color{Maroon}KoralW} cross sections: 
%Born and ISR level, 
%{\Color{Maroon}KoralW} ${\Color{Magenta}4f}$ correction,
%{\Color{Black}YFSWW3} ${\cal O}(\alpha)$ NL correction, at $200\,GeV$, 
%no cuts. The 
%{\Color{Red}last digits in $(\cdots)$} correspond to the statistical errors.}
%\label{YK-nocuts}
%\end{figure}
}
{\Color{Blue}  
{\small
\begin{table}[t]
\begin{center}
{\Color{Orange}$WW/4f$ Cross Section}\nl
\begin{tabular}{||c|c||c|c||c|c||c||}
\hline\hline
\multicolumn{2}{||c||}{\bf {\Color{Red}WITH CUTS}}  & 
\multicolumn{2}{c||}{$\sigma_{{\Color{Red}WW}}\,[fb]$} &
\multicolumn{2}{c||}{$\delta_{{\Color{Magenta}4f}}\,[\%]$} &
\raisebox{-1.5ex}[0cm][0cm]{$\delta_{{\Color{Red}WW}}^{NL}\,[\%]$} \\
%\hline
\cline{1-6}{\Color{Orange}
Final state} & {\Color{Orange}Program} & {\Color{Orange}Born} & {\Color{Orange}ISR} & {\Color{Orange}Born} &  {\Color{Orange}ISR} & \\
\hline\hline
& 
{\Color{Black}YFSWW3} & 
$210.938 \,(16)$ &
$196.205 \,(09)$ &
--- & --- &
$-1.93 \,(4)$ \\
{\Color{Green}$\nu_{\mu}\mu^+\tau^-\bar{\nu}_{\tau}$} &
{\Color{Maroon}KoralW} & 
$210.911 \,(26)$ &
$196.174 \,(21)$ &
${\Color{Magenta}0.041} $ &
${\Color{Magenta}0.044} $ & 
--- \\
\cline{2-4}
&
(Y$-$K)/Y &
$ 0.01 \,(1)\% $ &
$ 0.02 \,(1)\% $ &
--- & --- & --- \\
\hline\hline
 & 
{\Color{Black}YFSWW3} & 
$627.22 \,(5)$ &
$605.18 \,(3)$ &
--- & --- &
$-2.00 \, (4)$ \\
{\Color{PineGreen}$u\bar{d}\mu^-\bar{\nu}_{\mu}$} &
{\Color{Maroon}KoralW} & 
$627.13 \,(8)$ &
$605.03 \,(7)$ &
$ {\Color{Magenta}0.074} $ &
$ {\Color{Magenta}0.074} $ & 
--- \\
\cline{2-4}
&
(Y$-$K)/Y &
$ 0.01 \,(1)\% $ &
$ 0.02 \,(1)\% $ &
--- & --- & --- \\
\hline\hline
 & 
{\Color{Black}YFSWW3} & 
$1863.60 \, (15)$ &
$1865.00 \, (09)$ &
--- & --- &
$-2.06 \,(4)$ \\
{\Color{Brown}$u\bar{d} s\bar{c} $} &
{\Color{Maroon}KoralW} & 
$1863.07 \, (25) $ &
$1864.62 \, (21)  $ &
$ {\Color{Magenta}0.065} $ &
$ {\Color{Magenta}0.064} $ & 
--- \\
\cline{2-4}
&
(Y$-$K)/Y  &
$ 0.03 \,(2)\% $ &
$ 0.02 \,(1)\%$ &
--- & --- & --- \\
\hline\hline
\end{tabular}
%\caption{\sf
%The total $WW$ cross sections from YFSWW3 and KoralW 
%at the Born and ISR level, the $4f$ corrections from KoralW and
%${\cal O}(\alpha)$ NL correction from YFSWW3 at $E_{CM} = 200\,GeV$, 
%with cuts. 
%The numbers in parentheses are statistical
%errors of the results corresponding to last digits.
%}
%\label{tab:YR-xtot-withcuts}
%%%%%%%%%%%%%%%%%%%%%%%%%%%%%%%%%%%%%%%%%%%%%%%%%%%%%%%%%%%%%%%%%%%%%%%%%%%%
\end{center}
\caption{\Color{Orange}Total ${\Color{Red}WW}$  {\Color{Black}YFSWW3} and {\Color{Maroon}KoralW} cross sections: 
Born and ISR level, 
{\Color{Maroon}KoralW} ${\Color{Magenta}4f}$ correction, {\Color{Black}YFSWW3}
${\cal O}(\alpha)$ NL correction, at $200\,GeV$, with cuts. The
{\Color{Red}last digits in $(\cdots)$} correspond to the statistical errors.}
\label{YK-cuts}
\end{table}
}
%\begin{figure}[b!]
%\begin{center}
%%{\Color{Orange} WW/4f Cross Section}\nl
%%
%\setlength{\unitlength}{1mm}
%\begin{picture}(120,90)
%\put(45,60){\Color{Orange} $WW/4f$ Cross Section}
%%%\put(0,0){\framebox( 65,60){ }}
%\put(10,-40){\makebox(10,-40)[lb]{
%\epsfig{file=tabyvk-xtot1.eps,width=100mm,height=92mm}
%}}
%\end{picture}
%\end{center}
%\caption{\small Total ${\Color{Red}WW}$  {\Color{Black}YFSWW3} and {\Color{Maroon}KoralW} cross sections: 
%Born and ISR level, 
%{\Color{Maroon}KoralW} ${\Color{Magenta}4f}$ correction, {\Color{Black}YFSWW3}
%${\cal O}(\alpha)$ NL correction, at $200\,GeV$, with cuts. The
%{\Color{Red}last digits in $(\cdots)$} correspond to the statistical errors.}
%\label{YK-cuts}
%\end{figure}
}
\par
\section{Results-YFSZZ}
In this section we show the results of our comparison with ZZTO
for the total $ZZ$ pair signal processes as carried out in
Ref.~\cite{lep2YR:2000}. {\Color{Maroon}In that same set of comparisons,
ZZTO was also compared with the results of GENTLE by the authors
in Refs.~\cite{bard:2000}}. In this way, a cross check was made
on all three calculations.\par

{\Color{Blue}Specifically, we show in Table~\ref{zzschemes} the
ZZ signal cross section at 188.6 GeV as predicted by YFSZZ and ZZTO
for the case of no cuts as defined in Ref.~\cite{lep2YR:2000}.}
For ZZTO, results are shown for 
two schemes, {\Color{Red}the $G_\mu$ and $\alpha$
schemes~\cite{pass:2000}}. The agreement between
{\Color{Blue}  
%\end{center}
\let\sstl=\scriptscriptstyle
%%%=================local=macros=========================
\def\Was{W\c as}
\def\Order#1{${\cal O}(#1$)}
\def\Ordpr#1{${\cal O}(#1)_{prag}$}
\def\bbe{\bar{\beta}}
\def\tbe{\tilde{\beta}}
\def\tal{\tilde{\alpha}}
\def\tom{\tilde{\omega}}
\def\half{ {1\over 2} }
\def\alf1{ {\alpha\over\pi} }

\def\Oaz{${\cal O}(\alpha^0)$}
\def\Oaf{${\cal O}(\alpha^1)$}
\def\Oas{${\cal O}(\alpha^2)$}
%%%======================================================
% ============ begin table 1 ===============
{\small
%\begin{table}[hp]\centering
\begin{table}[t!]
\begin{center}
{\Color{Orange} Comparison with} {\Color{Magenta}ZZTO}
\begin{tabular}{|c|c|c|c|}
\hline
%% & & &\\
channel & {\tt YFSZZ} & {\tt {\Color{Magenta}ZZTO}} ${\Color{Red}G_F}$-scheme & {\tt {\Color{Magenta}ZZTO}} ${\Color{Red}\alpha}$-scheme \\
%% & & &\\
\hline
%% & & &\\
${\Color{Red}qqqq}$            &    294.6794(490) &  298.4411(60)  &    294.5715(59)   \\
${\Color{Green}qq\nu\nu}$        &    175.4404(302) &  175.5622(35)  &    174.9855(35)   \\
${\Color{Brown}qq{\rm ll}}$      &     88.1805(134) &   88.7146(18)  &     87.9881(18)   \\
${\Color{Orange}{\rm ll}\nu\nu}$  &     26.2530(463) &   26.0940(5)   &     26.1342(5)    \\
${\Color{Maroon}{\rm llll}}$      &      6.5983(15)  &    6.5929(1)   &      6.5706(1)    \\
${\Color{Black}\nu\nu\nu\nu}$    &     26.1080(71)  &   25.8192(5)   &     25.9868(5)    \\
total             &    617.2596(755) &  621.2241(124) &    616.2366(123)  \\
%% & & &\\
\hline
\end{tabular}
\vspace*{3mm}
%\caption[]{Comparison for the NC02 cross-section between {\tt YFSZZ} and
%{\tt ZZTO} at $\sqrt{s}= 188.6\,$GeV. The cross sections are in fb.
%\label{zzschemes}}
%\end{table}
%--
%}
\end{center}
\caption[]{ \Color{Orange} NC02 cross sections, {\tt {\Color{Blue}YFSZZ}} vs {\tt {\Color{Magenta}ZZTO}}, $188.6\,$GeV, in fb. The statistical
errors correspond to the last 
digits in
{\Color{PineGreen}$(~)$.}}
\label{zzschemes}
%\end{center}
\end{table}
}
%\end{picture}
}
\noindent the programs in this comparison and between the programs 
in the other related comparisons
carried out in Ref.~\cite{lep2YR:2000} show that {\Color{Red}the TU for the
respective NC02 signal process is 2\% at the respective LEP2 energies.}\par
\section{Conclusions}
{\Color{Blue}
  We are currently at an exciting point in the tests of the {\Color{Red}EW Theory}
in {\Color{Green}gauge boson physics}. The {\Color{Black}WW pair production} is an important
aspect of these tests. The radiative corrections which we realize
in YFSWW3-1.14 play a significant role in these tests as follows:
  \begin{itemize}
  \item
    {\Color{Black}Mass} distributions: these are affected by {\Color{Magenta}FSR}, yielding {\Color{Magenta}peak position} and {\Color{Magenta}height} shifts 
  \item
    {\Color{Black}W} Angular distributions: these are affected by {\Color{Magenta}LL} {\Color{Red}and} {\Color{Magenta}NL} corrections
  \item
    {\Color{Black}$\ell$} Angular distributions: these are affected by {\Color{Magenta}LL} {\Color{Red}and} {\Color{Magenta}NL} corrections
  \item
    {\Color{Black}Photon} Angular distributions: these are affected by {\Color{Magenta}LL} {\Color{Red}and} {\Color{Magenta}NL} corrections
  \item 
    {\Color{Black}Photon} Energy distributions: these are affected by {\Color{Magenta}LL} corrections
  \item
    {\Color{Black}Normalisation}: this is affected by {\Color{Magenta}LL} {\Color{Red}AND} {\Color{Magenta}NL} corrections;\\
    the current {\Color{Green}200 GeV} {\Color{Magenta}TU} is 0.4\% from the
\{{\tt \Color{Maroon}YFSWW3/RacoonWW}\} results.
  \end{itemize}
}
\par
Concerning our results on calculating the $4f$ background to {\tt YFSWW3}-1.14
using KoralW-1.42, we have shown the following:
{\bf\Color{Blue}
\begin{itemize}
\item
  Two different combinations of YFSWW3 and KoralW-1.42 cross sections reach the total precision {\Color{Red}${\cal O}(\frac{\alpha}{\pi}\frac{\Gamma_W}{M_W})$}.
  \item
   The size of the {\Color{Red}$4f$} correction to {\Color{Red}YFSWW3-1.14} is {\Color{Black}$\lesssim 0.1\%$}, as expected.
  \item 
   The future extension to a single platform is possible.
  \end{itemize}
}

{\bf\Color{Maroon}It follows that
YFSWW3$/$KoralW is a {\Color{Red}complete MC event generator} solution for precision
WW/4f production at LEP2 ( and LC's).\par
}
{\Color{Blue}
From our studies of the NC02 signal process we conclude 
that YFSZZ, a multiple photon {\Color{Red}MC event generator} for 
NC02 with $\bar\beta_0$ level LPA
YFS exponentiation (EEX), is 
tested in the LEP2 MC Workshop vs ZZTO and GENTLE to {\Color{Black}$ 2\%$
TU}. An
upgrade to higher precision is possible {\Color{Red}but is not needed, apparently}?
}

\Acknowledgments

Two of us (S.J. and B.F.L.W.) acknowledge the
kind hospitality of {\Color{Green}Prof. G. Altarelli and the CERN Theory 
Division} while this work was being completed. 
Three of us (B.F.L.W., W.P. and S.J.) 
acknowledge the support of {\Color{Green}Prof.~D.~Schlatter and Prof. D. Plane
and of the ALEPH, DELPHI, L3 and OPAL Collaborations}
in the final stages of this work. One of us (S.J.) is thankful
for the kind support of the {\Color{Green}DESY Directorate}
and one of us (Z.W.) acknowledges the support of the 
{\Color{Green}L3 Group of ETH Zurich} 
during the time this work was performed. All us thank the
members of the LEP2 MC Workshop for valuable interactions and
stimulation during the course of this work. The authors especially thank
{\Color{Magenta}Profs. A. Denner, S. Dittmaier and F. Jegerlehner and Drs.
M. Roth and D. Wackeroth} for useful discussions and interactions.

\end{document}